\documentclass[conference]{IEEEtran}
\IEEEoverridecommandlockouts
\usepackage{amsmath,amssymb,amsfonts}
\usepackage{algorithmic}
\usepackage{graphicx}
\usepackage{textcomp}
\usepackage{xcolor}
\usepackage{float}
\usepackage{caption}

\usepackage[justification=centering]{caption}

\def\BibTeX{{\rm B\kern-.05em{\sc i\kern-.025em b}\kern-.08em
    T\kern-.1667em\lower.7ex\hbox{E}\kern-.125emX}}

\begin{document}

\title{{A SUMO-Based Digital Twin for Evaluation of Conventional and Electric Vehicle Networks}
}

\author{Haomiaomiao Wang, Conor Fennell, Swati Poojary and Mingming Liu \thanks{H. Wang and M. Liu are with the Insight Research Ireland Centre for Data Analytics at Dublin City University, Ireland. C. Fennell and S. Poojary are with the School of Computer Science and Statistics, Trinity College Dublin. \textit{Corresponding author: Mingming Liu. Email: {\tt mingming.liu@dcu.ie}.}}}

\maketitle

\begin{abstract}
Digital twins are increasingly applied in transportation modeling to replicate real-world traffic dynamics and evaluate mobility and energy efficiency. This study presents a SUMO-based digital twin that simulates mixed ICEV–EV traffic on a major motorway segment, leveraging multi-sensor data fusion from inductive loops, cameras, GPS probes, and toll records. The model is validated under both complete and partial information scenarios, achieving 93.1\% accuracy in average speed estimation and 97.1\% in average trip length estimation. Statistical metrics, including KL Divergence and Wasserstein Distance, demonstrate strong alignment between simulated and observed traffic patterns. Furthermore, CO$_2$ emissions were overestimated by only 0.8–2.4\%, and EV power consumption underestimated by 1.0–5.4\%, highlighting the model’s robustness even with incomplete vehicle classification information.
\end{abstract}


\section{Introduction}

Digital twin technology has gained prominence in transportation research by providing virtual replicas of traffic systems before real-world deployment \cite{kuvsic2023digital}. Existing studies have demonstrated the efficacy of leveraging digital twins for predicting traffic flows and optimizing transportation networks \cite{argota2022getting}, \cite{nag2025exploring}. These twins are particularly valuable in evaluating emerging technologies such as electric vehicles (EVs), where data scarcity and the complexity of mixed-fleet interactions pose challenges for traditional modelling approaches \cite{sumitkumar2024shared}.

Despite the rapid adoption of EVs and their potential to significantly reduce emissions \cite{ke2017assessing}, there is limited research on evaluating EV energy consumption within digital twin frameworks. Furthermore, the increasing level of EVs on road networks also creates complex interactions with internal combustion engine vehicles (ICEVs) through traffic flow \cite{li2024experimental}, where they interact in non-linear and dynamic ways in mixed traffic environments \cite{he2024introducing}. How to accurately capture these interactions remains a significant challenge. Existing research often relies on oversimplified assumptions, which limits the ability to fully understand and optimize the performance of both EV and ICEV networks \cite{javid2017comprehensive}.

\begin{figure}[htbp]
    \centering
    \includegraphics[width=\linewidth]{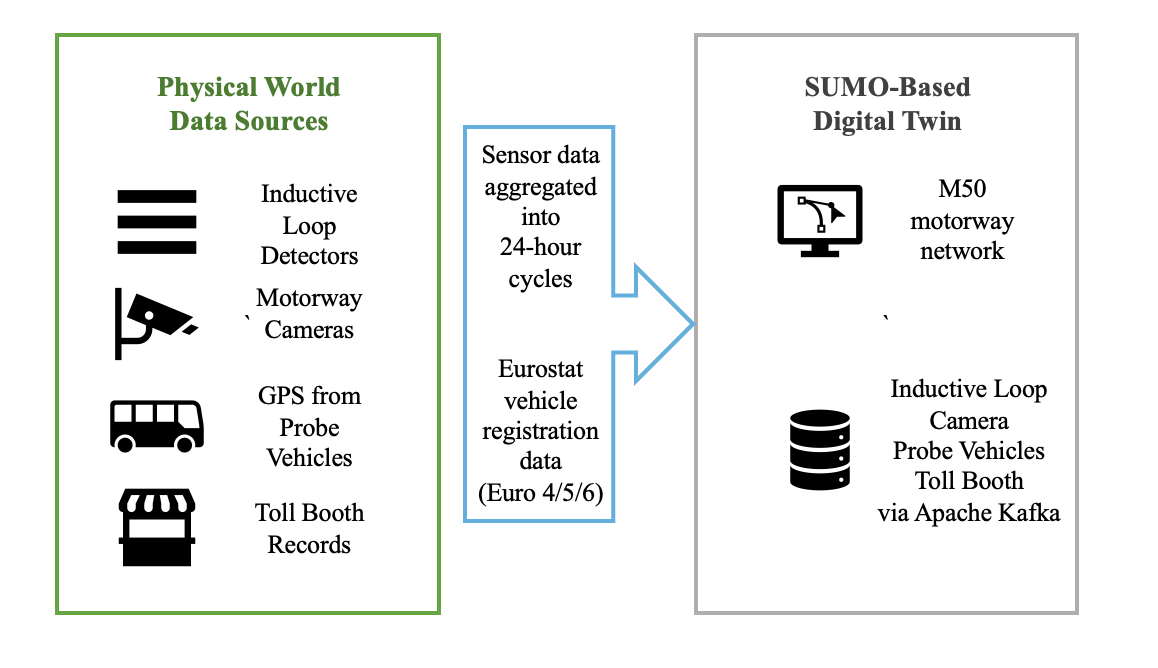}
    \caption{The SUMO-based digital twin framework.}
    \label{dt}
\end{figure}

This paper aims to take the first step towards addressing this challenge by proposing a digital twin-based method that focuses on estimating the energy consumption of EVs and emissions from ICEVs under realistic mixed-traffic conditions. A validated digital twin framework using the Simulation of Urban Mobility (SUMO) platform integrates heterogeneous sensor data from physical infrastructure, including inductive loops, cameras, probe vehicles, and toll booth records, as shown in Fig. \ref{dt}. A representative section of Dublin’s M50 motorway was selected for analysis in this work thanks to its rich traffic data sources. The digital twin replicates real-world traffic behaviour and enables systematic evaluation of energy and emissions under varying EV penetration scenarios. The schematic diagram of the digital twin workflow also reflects the integration of Eurostat registration data and the Apache Kafka streaming pipeline, following the methodological principles outlined in prior thesis work \cite{fennellCommunicationArchitectureTransportation2022}, \cite{poojaryFrameworkBuildingDigital}.

Our contributions are threefold: (1) developing and validating a SUMO-based digital twin simulation capable of accurately replicating real-world traffic interactions; (2) introducing a refined methodology for assessing vehicle-specific energy and emissions profiles under dynamic conditions; (3) offering quantitative insights into how EV-ICEV interactions influence overall transportation efficiency and environmental outcomes.

The remainder of the paper is structured as follows: Section II reviews related work on digital twins and mixed-traffic modelling. Section III introduces the methodology, including digital twin validation and energy modelling for both ICEVs and EVs. Section IV describes the simulation setup and data collection process. Section V presents the validation results and performance analysis. Finally, Section VI concludes the paper and discusses potential directions for future research.

\section{Related Works}

The use of digital twins in transportation has advanced from simple models to real-time systems that require continuous validation against real-world data to ensure accuracy \cite{wu2025digital, pikner2024autonomous}. A key application is assessing vehicle performance in mixed-traffic scenarios composed of both EVs and ICEVs \cite{nag2025exploring}. Recent research confirms that EVs are generally more energy-efficient than ICEVs, particularly in stop-and-go urban traffic where they benefit from regenerative braking \cite{dong2022comparative, ahani2023optimization, yan2023privacy}. In contrast, ICEVs perform more efficiently during highway driving but produce higher emissions in congestion \cite{fiori2019effect}. These operational differences are captured in microscopic cost function models, which account for factors like engine load for ICEVs and auxiliary power use for EVs \cite{gahlaut2024comprehensive, liu2015intelligent}. However, there is a lack of research that uses validated digital twins to systematically compare these vehicle types in mixed-traffic environments \cite{li2023survey}. This limitation highlights the need for a framework that can accurately evaluate how overall network efficiency changes with varying levels of EV penetration.

\section{Methodology}

\subsection{Rationale for a SUMO-Based Digital Twin}

This study employs a SUMO-based digital twin methodology for real-world traffic simulation, including vehicle interactions, lane-changing behaviour, and congestion patterns. To enhance simulation fidelity, our framework integrates multi-sensor data fusion with real-time traffic counts and mobility datasets for cross-validation. The chosen simulation site is a 7 km, 4-lane section of the M50 motorway with two major interchanges in Dublin, Ireland. This network leverages comprehensive and well-documented traffic flow data recorded in Ireland from 2012 to 2019, reorganised into a representative 24-hour cycle \cite{gueriau2020quantifying}. This dataset provides a robust foundation for validation and analysis of mixed-traffic conditions.

In our framework, both the physical and digital twins are developed using SUMO. The physical twin refers to a high-fidelity simulation that incorporates complete information from the sensor-derived dataset and serves as a surrogate for the real-world system \cite{gueriau2020quantifying}. The digital twin, on the other hand, is constructed under more realistic constraints with partial information and probabilistic vehicle classification. Using SUMO for both models ensures consistency in traffic dynamics, network structure, and vehicle behaviours, and enables systematic validation and scenario testing. This approach is consistent with recent digital twin literature in transport systems \cite{fennellCommunicationArchitectureTransportation2022, poojaryFrameworkBuildingDigital}.

The SUMO-based simulation leverages this collected data to produce detailed visualisations of traffic scenarios and enable controlled testing of different traffic compositions, which supports urban and interurban traffic analyses and prediction.

\subsection{Digital Twin Validation and Calibration}

To ensure the digital twin accurately replicates real-world traffic, this study employs a multi-metric validation approach to evaluating both fundamental vehicle properties and statistical distribution alignment. Beyond basic traffic validation, distributional metrics are incorporated to capture the differences between real-world and simulated traffic distributions, addressing limitations in traditional validation approaches. The statistical divergence metrics used in this framework include: 
\begin{itemize}
    \item Kullback-Leibler (KL) Divergence – Measures how one probability distribution deviates from another \cite{amari2018information}.
    \item Jensen-Shannon (JS) Divergence – A symmetric extension of KL Divergence for balanced similarity assessment \cite{nguyen2015non}.
    \item Wasserstein Distance – Evaluates the effort required to transform one distribution into another \cite{panaretos2019statistical}.
    \item Bhattacharyya Distance – Measures overlap between two probability distributions \cite{andersen2022evaluation}.
\end{itemize}

By iteratively applying these statistical measures, the digital twin is fine-tuned to improve traffic flow characteristics, including speed distributions and route accuracy. Unlike conventional validation methods that rely solely on error-based metrics, this approach captures the probabilistic nature of traffic variations \cite{gomes2018methodology}. Through systematic reduction of distributional divergence, the finalised SUMO-based digital twin achieves a high degree of similarity to real-world traffic behaviour. This approach strengthens the credibility of the digital twin as a tool for evaluating different traffic compositions, particularly in mixed ICEV-EV scenarios. 

\subsection{EV and ICEV Performance Modelling}
SUMO supports flexible configurations of ICEV and EV proportions, enabling comparative performance evaluations across various fleet compositions. This enhances the realism of energy efficiency assessments in mixed-traffic scenarios, addressing a significant limitation in current digital twin models. To best quantify CO$_2$ emissions for ICEVs and energy consumption for EVs, cost functions are applied based on the assumptions outlined in \cite{liu2015distributed}. In particular, ICEV CO$_2$ emissions per unit distance are modelled based on the following average-speed model:
\begin{equation}
E_{CO_2} = k \left( \frac{a + b s_i + c s^2 + d s^3 + e s^4 + f s^5 + g s^6}{s} \right)\label{icev}
\end{equation}
where \( s \) represents the vehicle speed, and the coefficients \( a, b, c, d, e, f, g \) are empirically derived coefficients that define CO$_2$ emissions standards across vehicle classes \cite{liu2015distributed}.  

Power consumption per unit distance for EVs is expressed as:
\begin{equation}
\frac{P_{cons}}{v} = \frac{\alpha_0}{v} + \alpha_1  + \alpha_2 v + \alpha_3 v^2 \label{ev}
\end{equation}
where \( v \) represents the vehicle speed, and the parameters \( \alpha_0, \alpha_1, \alpha_2, \alpha_3 \) are empirically derived coefficients capturing the effects of aerodynamic drag, rolling resistance, and drivetrain losses on EV power consumption under varying speed conditions \cite{liu2015distributed}, \cite{liu2015intelligent}.


\section{Experiment Setup}

\begin{figure*}[htbp]
    \centering
    \includegraphics[width=\linewidth]{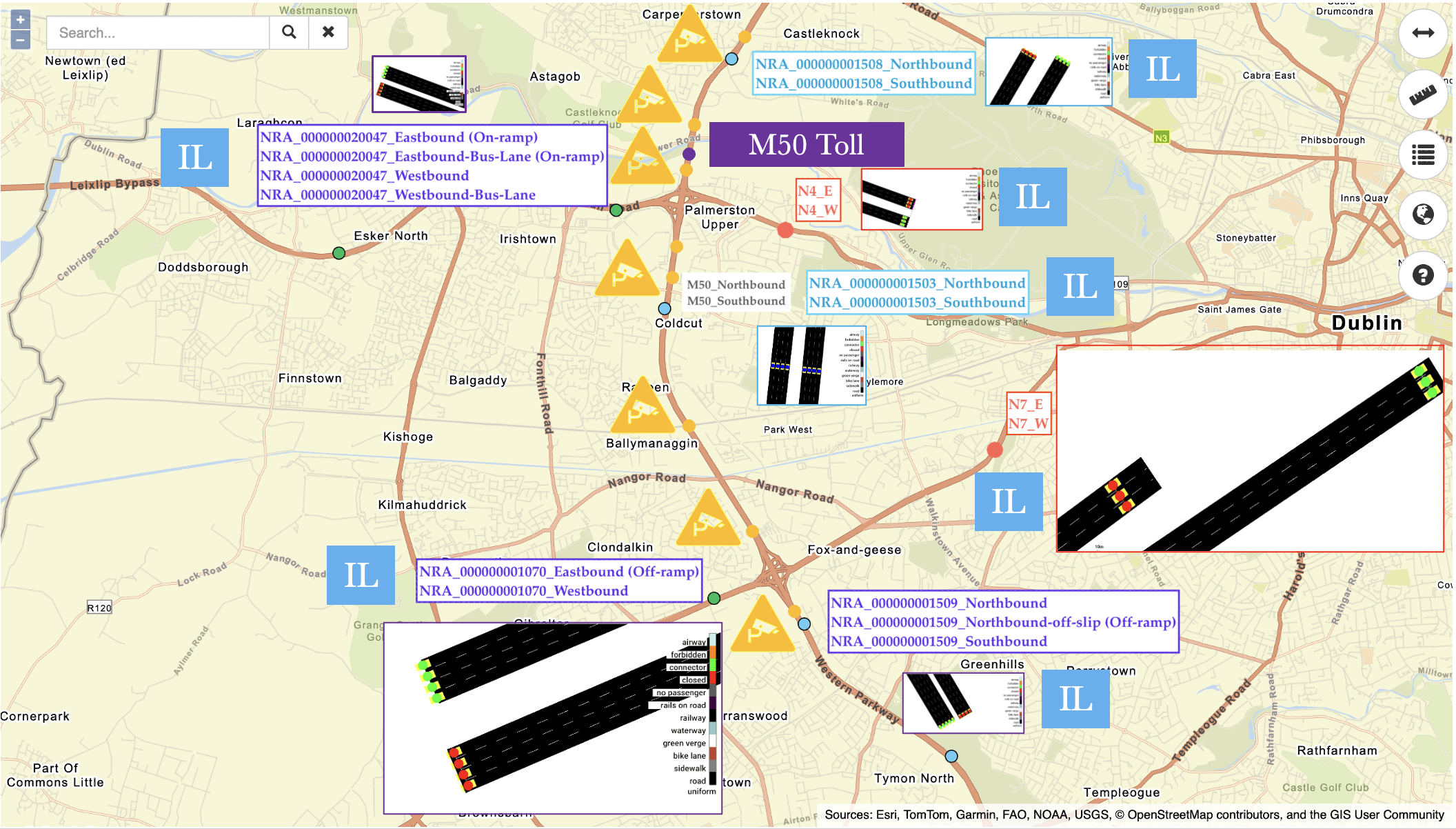}
    \caption{Simulated M50 Motorway Network in Dublin, Ireland.}
    \label{m50}
\end{figure*}

Building upon the methodological framework, this study deploys a SUMO-based simulation on Dublin's M50 motorway, specifically at its intersections with the N4 and N7 national roads. The segment is extensively instrumented with sensors, including 7 sets of inductive loop detectors, 7 motorway cameras, 1 toll bridge, and multiple probe vehicles equipped with GPS tracking, as illustrated in Fig. \ref{m50}. These sensors provide comprehensive data on traffic volumes, vehicle speeds, lane positions, and congestion patterns, facilitating accurate calibration of the simulation.

Given the infeasibility of direct sensor identification of ICEV emission standards, Eurostat vehicle registration data was used to estimate ICEV Euro-class distribution, with assigned probabilities of 14.8\% Euro 4, 21.8\% Euro 5, and 63.4\% Euro 6 \cite{eurostat2024passenger}. Corresponding CO$_2$ emission factors, listed in Table \ref{tab:co2_emission_factors}, are integrated into the emission model in \eqref{icev}, modelling emissions for different types of ICEVs  \cite{boulter2009emission}.

\begin{table}[htbp]
\caption{CO$_2$ Emission Factors}
\footnotesize 
\setlength{\tabcolsep}{3.5pt} 
\renewcommand{\arraystretch}{1.1} 
\begin{tabular}{|c|c|c|c|c|c|c|}
\hline
\textbf{Class} & \textbf{a} & \textbf{b} & \textbf{c} & \textbf{d} & \textbf{e, f, g} & \textbf{k} \\
\hline
Euro 4 & 3.7473E+03 & 1.5599E+02 & -8.5270E-01 & 1.0318E-02 & 0 & 1 \\
Euro 5 & 3.7473E+03 & 1.2877E+02 & -8.5270E-01 & 1.0318E-02 & 0 & 1 \\
Euro 6 & 3.7473E+03 & 1.0571E+02 & -8.5270E-01 & 1.0318E-02 & 0 & 1 \\
\hline
\end{tabular}
\label{tab:co2_emission_factors}
\end{table}

To accurately model the variability in EV energy consumption within the digital twin, parameters ($\alpha_0, \alpha_1, \alpha_2, \alpha_3$) were determined probabilistically, reflecting realistic fleet heterogeneity. Four sources of energy loss were explicitly modelled as follows:

\begin{itemize}
    \item \text{Ancillary system power} ($\alpha_0$): Representing constant auxiliary energy demand, drawn from a uniform distribution:
    \[
    \alpha_0 = 0.2 + U(0, 2.0)
    \]
    \item \text{Drivetrain losses} ($\alpha_1$): Assumed constant, reflecting consistent drivetrain efficiency across EVs:
    \[
    \alpha_1 = 5\times10^{-4}
    \] 
    \item \text{Rolling resistance} ($\alpha_2$): Rolling resistance was proportional to total vehicle mass, including passenger load ($n_{\text{pass}}$, randomly between 1 and 4 passengers with average passenger weight of 80 kg):
    \[
    \alpha_2 = 0.0293 + 0.05\left(\frac{1235 + 80 \times n_{\text{pass}}}{1235}\right)
    \]
    \item \text{Aerodynamic power losses} ($\alpha_3$): Aerodynamic losses depend significantly on vehicle design and speed, modelled as:
    \[
    \alpha_3 = P_{\text{aer}} + 4\times10^{-6}
    \]
    \[
    \quad P_{\text{aer}} \sim \mathcal{N}(3.12\times10^{-5},\, (5\times10^{-6})^2)
    \]
\end{itemize}

This parametrization approach ensures a robust and realistic representation of EV energy consumption, capturing practical variations within a mixed-vehicle fleet for digital twin simulations.

To systematically assess the digital twin’s robustness, three clearly defined experimental scenarios were considered. The Complete Information Digital Twin (CIDT) scenario assumes full knowledge of vehicle parameters, including precise ICEV emission classes, EV energy consumption details, vehicle speeds, and lane-changing behaviours, representing ideal data collection conditions. The Partial Information Digital Twin (PIDT) scenario introduces uncertainty by probabilistically inferring vehicle classifications and traffic parameters, reflecting realistic sensor limitations. Finally, the Physical Entity Data scenario provides empirical sensor measurements from the M50 motorway, offering a ground-truth baseline for validating the simulation results. A Python-based validation script quantified scenario accuracy through systematic comparisons of CIDT, PIDT, and empirical data, analysing vehicle speed distributions, total CO$_2$ emissions for ICEVs, and energy consumption for EVs.

\section{Results}

\subsection{Validation of the Digital Twin Accuracy}

\begin{figure}[htbp]
    \centering
    \includegraphics[width=\linewidth]{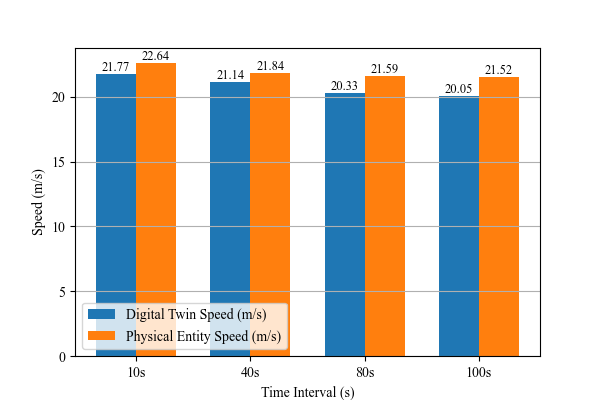}
    \caption{Comparison of average speed for vehicles in both digital twin and physical entity environments.}
    \label{speed}
\end{figure}

\begin{figure}[htbp]
    \centering
    \includegraphics[width=\linewidth]{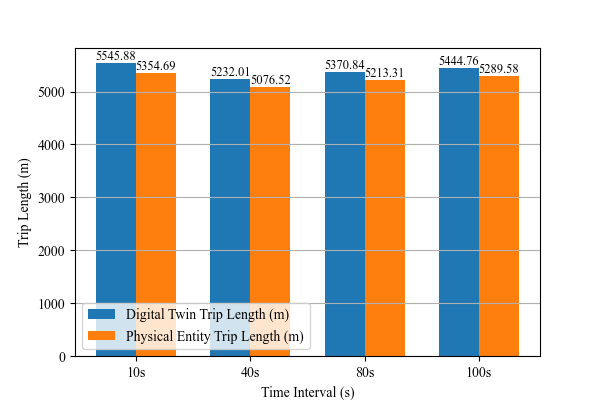}
    \caption{Comparison of average trip distance for vehicles in both digital twin and physical entity environments.}
    \label{length}
\end{figure}

\begin{figure}[htbp]
	\centering
	\includegraphics[width=\linewidth]{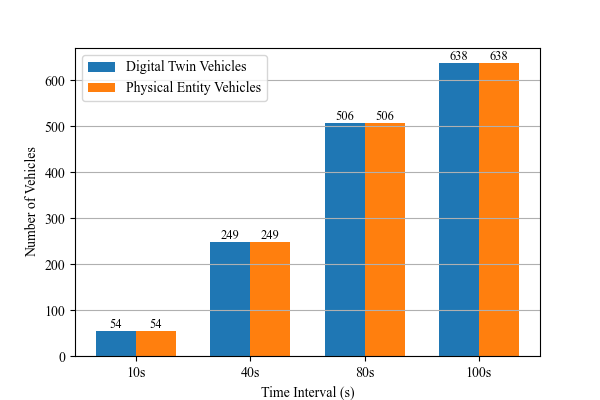}
	\caption{Comparison of the total count of vehicles in both digital twin and physical entity environments.}
	\label{count}
\end{figure}

To evaluate the accuracy of the digital twin, key metrics including average speed, average trip length, and the total number of vehicles, were compared between simulation and real-world observations. Vehicles were emitted within time intervals of 10s, 40s, 80s, and 100s; however, each vehicle completed its route along the M50 motorway irrespective of the emission interval. Simulated average speeds achieved 93.1\% accuracy compared to real-world data. However, the digital twin consistently underestimates average vehicle speeds across all time intervals in Fig. \ref{speed}. On the other hand, the digital twin achieved a 97.1\% accuracy in estimating average trip lengths, although it consistently overestimated them, as shown in Fig. \ref{length}. These systematic deviations may be attributed to specific SUMO parameterisations, particularly the conservative settings in the default lane-changing model and the IDM car-following model, which can lead to more cautious driving behaviour and suboptimal lane utilization \cite{xiao2020safety}. In contrast, vehicle count validation in Fig. \ref{count} showed that the simulation accurately replicated the total observed traffic volume across all intervals, confirming the effectiveness of the vehicle generation process in matching empirical inflow rates. 

To further quantify digital twin accuracy, four statistical metrics were calculated to measure the similarity of average speed distribution. Measuring similarity in speed distributions is particularly important, as speed inherently integrates both trip durations and trip lengths, capturing comprehensive travel behaviours. Accurate representation of speed distribution directly impacts subsequent emissions and energy consumption calculations, since CO$_2$ emissions for ICEVs and energy usage for EVs strongly depend on the precise vehicle speed profiles throughout their trips. Therefore, validating speed distribution similarity ensures reliable downstream evaluations of environmental impacts and energy efficiency.

\begin{figure}[htbp]
    \centering
    \includegraphics[width=\linewidth]{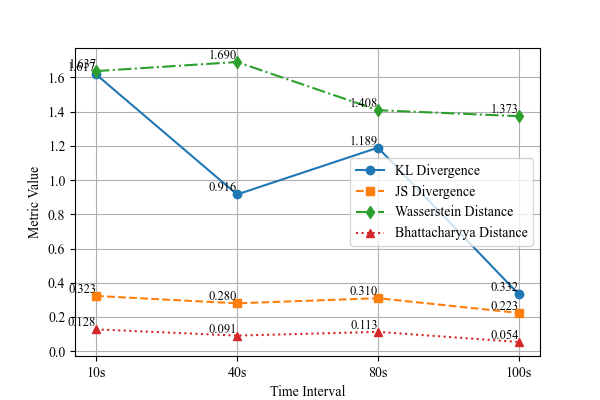}
    \caption{Comparison of different statistical metrics over different time intervals.}
    \label{metric}
\end{figure}

These metrics offer complementary perspectives on distributional alignment: KL and JS Divergence assess divergence between distributions, Wasserstein Distance measures the cost to align distributions, and Bhattacharyya Distance captures their overlap. A considerable decrease in both KL and JS Divergences indicates that the digital twin more accurately replicates real-world driving behaviours as traffic volume increases in Fig. \ref{metric}. Likewise, the declining trends in Wasserstein Distance and Bhattacharyya Distance further confirm that increased traffic volume enhance alignment between the digital twin and real-world conditions. 

\subsection{CO$_2$ Emissions and Power Consumption under Different EV Penetration Levels}
\begin{figure}[htbp]
	\centering
	\includegraphics[width=\linewidth]{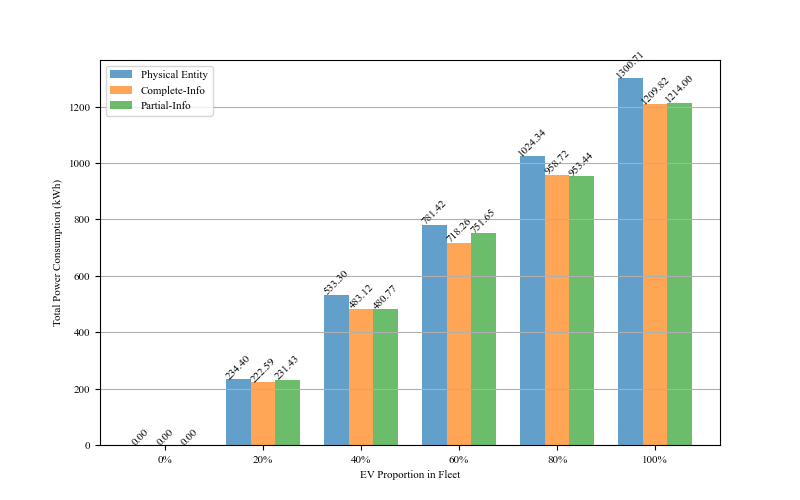}
	\caption{Comparison of the total CO$_2$ emissions of ICEVs by different penetration levels in three settings.}
	\label{cost}
\end{figure}

\begin{figure}[htbp]
	\centering
	\includegraphics[width=\linewidth]{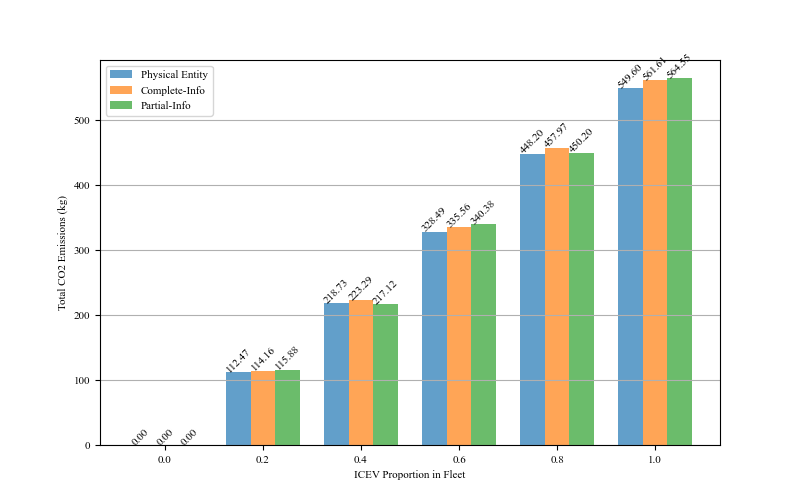}
	\caption{Comparison of the total energy consumptions of EVs by different penetration levels in three settings.}
	\label{energy}
\end{figure}

The CO$_2$ emissions and power consumption across various EV penetration levels evaluate the digital twin's capability under scenarios with complete and partial information, compared to the physical entity according to Fig. \ref{cost} and Fig. \ref{energy}. Total CO$_2$ emissions consistently increase with higher proportions of ICEVs in the fleet, with the digital twin slightly overestimating emissions by 0.8\% to 2.4\% across different fleet compositions. Notably, the discrepancy diminishes as the ICEV proportions decreases, indicating higher digital twin accuracy in mixed or EV-dominated fleets. 

For power consumption, the digital twin underestimates energy usage by 1.0\% to 5.4\%, with the most significant difference occurring at mid-range EV penetration levels. However, the deviation remains relatively stable, confirming that the model provides a reliable representation of energy dynamics in mixed-traffic conditions. The observed error margins, 2.4\% for CO$_2$ emissions and 5.4\% for EV power consumption, are relatively small, especially considering the partial information scenario. In practical terms, these deviations fall well within acceptable bounds for urban-scale transportation modelling and policymaking, where uncertainty margins of $\pm$ 5–10\% are common in forecasting and planning \cite{bastani2012effect}, \cite{profillidis2018modeling}.

Overall, these findings demonstrate the SUMO-based digital twin effectively captures the general trends of emissions and power consumption under varying fleet compositions, despite minor discrepancies. These insights support the reliability and applicability of the digital twin for evaluating transportation strategies aimed at reducing emissions and enhancing energy efficiency in realistic, mixed-traffic conditions. A key outcome of our study is the demonstrated robustness of the digital twin under the PIDT scenario. Given that transport authorities often operate with fragmented or incomplete data sources, the strong performance of the PIDT underscores the practical applicability of our framework in real-world scenarios.

\section{Conclusion}

This paper has developed a SUMO-based digital twin to simulate mixed-traffic conditions, integrating multi-sensor data fusion for improved accuracy. The validation results have demonstrate an accuracy of 93.1\% for average speed and 97.1\% for average trip length using the proposed framework in a case study conducted on the Dublin’s M50 motorway. Distributional similarity metrics have been applied to further validate the framework's accuracy in modelling real-world traffic distributions. The framework was tested under both complete and partial information scenarios with results showing that CO$_2$ emissions overestimations of only 0.8\%–2.4\% and underestimations in power consumption of 1.0\%–5.4\%, demonstrating its robustness even with incomplete vehicle classification data. 

While the current emission and power consumption models are based on average-speed formulations, which offer computational efficiency, they inherently simplify vehicle dynamics. Future work will explore the integration of instantaneous or time-resolved models to better capture transient behaviours such as acceleration spikes, deceleration events, and regenerative braking effects in EVs. Additionally, we plan to conduct sensor-type sensitivity analyses to quantify the relative contribution of each data source to the overall digital twin accuracy, helping to prioritize future infrastructure investment. The framework will also be scaled to larger and more complex non-highway urban networks, where heterogeneity in vehicle behaviour and road structure may pose new challenges. These extensions aim to further improve the realism and policy relevance of the digital twin for broader deployment in intelligent transport systems.

\section*{Acknowledgment}
This work has emanated from research conducted with the financial support of Taighde Éireann — Research Ireland under Grant number \textit{SFI/12/RC/2289\_P2}  and co-funded by the European Regional Development Fund in collaboration with the Research Ireland Insight Centre for Data Analytics at Dublin City University.
\bibliographystyle{IEEEtran}
\bibliography{References.bib} 
\vspace{12pt}
\end{document}